\documentclass[journal]{IEEEtran}

 \usepackage[pdftex]{graphicx}

\title{Searchers Seeking: \\What Happens When you Frustrate Searchers? }
\author{Gareth Renaud\\School of Computing Science\\University of Glasgow}

\begin{document}

\maketitle


\begin{abstract}
People searching for information occasionally experience difficulties finding what they want
on the Web. This might happen if they cannot quite come up with the right search terms.
What do searchers do when this happens?
Intuitively one imagines that they will try a number of associated search terms
 to zero in on their intended search target. Certainly the provision of spelling suggestions and related search terms
assume that frustrated searchers will use these to implement this strategy. 
Is this assumption correct? What do people really do?

We ran an experiment
where we asked people to find some relevant links, but we prevented them from using the most obvious search terms,
which we termed ``taboo words". 
To make the experiment more interesting we also provided the traditional forms of assistance: spelling suggestions and related search suggestions.
We assigned participants using a magic square to get no assistance, one kind of assistance, or both. 
Forty eight people participated in the experiment. 

What emerged from the analysis was that when people are frustrated in their searching attempts, a minority soldier on, attempting to find other terms, but the majority will stick with their original query term and 
simply progress from page to page in a vain attempt to find something relevant. This confirms findings by other researchers about the difficulties of query re-formulation.  Our finding will serve to inform the developers of user interfaces to search engines, since it would be helpful
if we could find a better way of supporting frustrated searchers. 
 
\end{abstract}


\section{Introduction}
Searching for information is something most of us engage in, using search engines such as Google, Bing and Yahoo. When you can provide the right search term, you will be more likely to find what you're looking for. Sometimes, however, it is difficult to come up with the right search terms, and you have to try a variety of related search terms to home in on desired information.
The ``tip of the tongue'' experience \cite{Brown66}, where one
simply cannot come up with the right word to describe something, is vaguely related to this, since sufferers experience the same frustrations as those who can't quite come up with the right search term.

There is a need to start focusing on the user experience of information retrieval. Martzoukou \cite{Martzoukou} calls for this behaviour to be studied from all its multiple facets, including information need, which is the focus of this study. Hseih-Yee \cite{Hseih} also argues for the need to find out how the current Web environment supports or hinders information seeking. 

\begin{figure}[h]
\begin{center}
\includegraphics[width=8cm]{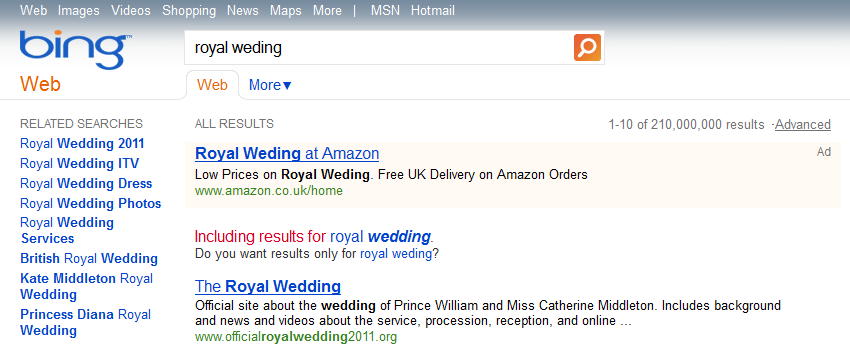}
\caption{Bing offers Spelling and Related Search Suggestions}
\label{fig_bing}
\end{center}
\end{figure}

\begin{figure}[h]
\begin{center}
\includegraphics[width=8cm]{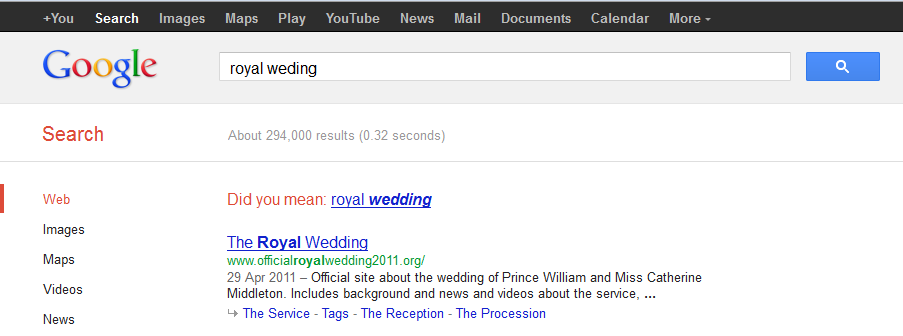}
\caption{Google Offers Spelling Suggestions, with related search suggestions at bottom of page}
\label{fig_google}
\end{center}
\end{figure}

\begin{figure}[h]
\begin{center}
\includegraphics[width=8cm]{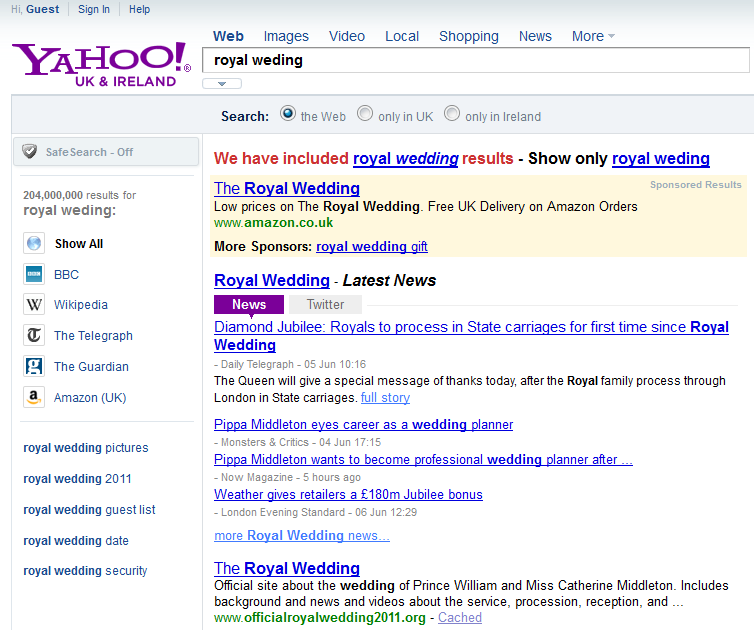}
\caption{Yahoo offers Spelling and Related Search Suggestions}
\label{fig_yahoo}
\end{center}
\end{figure}

Search engines commonly offer spelling suggestions and/or related search suggestions.
Figures  \ref{fig_bing}, \ref{fig_google}  and \ref{fig_yahoo} show the different approaches taken by the most popular search engine interfaces. Bing and Yahoo provide a list of related searches to the left of the search pane, but Yahoo places these below the filtering links which narrow down the search while Bing places them directly to the left. Google provides them at the bottom of the page. All engines, when they do suggest spelling, provide only one spelling suggestion. No doubt this is based on sophisticated data mining of submitted search terms, so that the best possible term can be presented. 

The following section reviews the literature in this area.

\section{Background}\label{backg}

Do web searchers need assistance, and does provided assistance help them to find what they are looking for? Wolfram {\em et al.} \cite{SpinkBrief}
have argued for this question to be addressed, so that, as the Web evolves at such a massive scale, we can start to understand better how to support Web searchers. 

\subsection{Is Assistance Needed?}
Sutcliffe {\em et al.} \cite{Sutcliffe} carried out an investigation into searcher performance and found that overall performance was poor. They observed searchers using the wrong search terms, amongst other poor strategies. They also found that searchers did not use provided assistance
if they had to engage in several interaction steps to do so.
Jansen {\em et al.} \cite{Jansen} analysed transaction logs and found that
relevance feedback was used very little. They also found that most searchers searched with one query only, and did not follow up with subsequent queries. They seldom viewed more than 2 pages. 
All of these studies suggest the need to provide as much assistance as possible without requiring action on the user's part. Spink {\em et al.}'s 
study confirms these findings \cite{Ozmultu}.  They studied over a million web searches and concluded that searching was a low art.
Indeed, Jansen and Spink \cite{JansenSpink2} determined that only 50\% of documents found by searchers were relevant.   

H\"{o}lscher and Strube \cite{Holscher} found that two kinds of expertise were needed to search effectively: web experience and domain knowledge.
The need for domain knowledge is even more important when one considers that Broder \cite{Broder} and Rose and Levinson \cite{Rose} determined that, of the three kinds of searches people engage in, informational search made up roughly half of all searches. The other two: navigational (give me the right url) and transactional (wanting to shop or perform some other activity) made up the other half, and did not, presumably require domain knowledge. 

In conclusion, it seems that searchers could well benefit from assistance. Clearly the main search engines have come to the same conclusion, as evidenced by the screenshots in the introduction.   

\subsection{What Kinds of Assistance?}
Jansen \cite{Jansen00} reports that users routinely mis-spell search terms, so it makes sense to offer spelling suggestions.
McCray {\em et al.} \cite{McCray} argue that offering suggestions to searchers will help them to re-formulate their queries, and in the end to find what they seek. 
They did not carry out an experiment to prove that this would work, however.
Kelly {\em et al.} \cite{Kelly09}, in their experiment, found no difference in performance when searchers were offered suggestions based on term relevance feedback, or user generated suggestions. They only compared two different kinds of suggestions, and did not have a control group who did not get any suggestions at all, so it is difficult to measure the performance impact of either type of suggestion as opposed to none at all.

Fonseca {\em et al.} \cite{Fonseca03} report that when they offered related search suggestions, and users clicked on them, in 92\% of cases users found something useful. 
However Anick {\em et al.} \cite{Anick03} were not able to prove an significant effect of offering terminological feedback to searchers. 
None of these studies were aimed at people who couldn't quite come up with the correct word to start their search. 

The rest of this paper is organised as follows.
Section \ref{options} provides details the design of the experiment.
and introduces the SCAMP tool which was used
to support this research. Section \ref{scamp} explains how SCAMP was configured to
carry out the experiment. 
Section \ref{results} presents
the results and Section \ref{conc} concludes.

\section{Investigation into Assistance Options}\label{options}

A study was conducted in order to 
 to investigate which aids {\em do} help web searchers. The experiment contrived to prevent searchers from using the most obvious terms in searching for news items, so as to emulate the situation where people couldn't initially formulate effective query terms. Research \cite{Ozmultu} suggests that
only one in five web searchers re-formulate queries, so this experiment
sought to compel users to do so, in order to ensure that
they looked for and used any assistance that was required. 
Three research questions were addressed by this study:

\begin{enumerate}
\item Which aid: spelling or related search suggestions, is most helpful to users who have difficulty finding what they want?

\item Where  on the screen should these suggestions be presented?

\item How many suggestions, of each type, should be offered? 

\end{enumerate}

\noindent The hypothesis being tested is:
\begin{quote}
{\em
There will be a direct relationship between the number of search aids provided and the number of
positive relevance judgements given by participants in an experiment where participants are asked
to find a topic without using a given set of words related to the search topic.
}
\end{quote}

\subsection{The SCAMP Tool}

Since search engine assistance was being tested it made sense to harness web services which are offered by some web search engines (eg. Bing, Digg). A generic tool, called SCAMP{\footnote{SCAMP is available from http://sourceforge.net/projects/puppyir/.}}, was developed to conduct this experiment. The need for SCAMP is emphasised by the findings of
Jansen and Spink \cite{JansenSpink} who found that findings from user-based research focused
on web searching with one particular engine cannot be applied to
searching with another engine so there is clearly a need for a tool which
supports experimental comparison.

SCAMP allows researchers to configure and run online web engine related experiments. SCAMP supports the researcher in configuring the experiment in a particular way to test configurations, assistance mechanisms, and timings. Furthermore, it allows the researcher to integrate 
questionnaires before and after tasks and for the entire experiment. 
The experiment, once built, can be activated so that participants can register, consent and participate. Everything is done remotely, and the researcher is able to monitor the progress of the experiment via SCAMP's researcher dashboard. A full description of SCAMP is provided here: \cite{RenaudG}. SCAMP is a generic tool, able to support a range of comparisons between search engine configurations. This particular experiment was an instance of the kind of experiment SCAMP can facilitate. 

\subsection{Implementation Decisions}

In order to answer the three research questions,  the following implementation decisions were made:
\begin{itemize}
\item SCAMP applies a latin square rotation to test the effects of spelling suggestions, related search suggestions, both or neither. 
A within-subjects design was implemented, so that participants would execute search tasks for each of the different conditions. 
\item The previous design decision impacted on the decision
of where to display the assistance. 
All the popular interfaces offer the spelling suggestion directly under the search box, and those that offer related search suggestions display them in a panel to the left. 
For this experimental design four conditions applied: {\em no} suggestions, {\em one} of the two, or {\em both}.
If one follows the lead of the major search engines  spelling suggestions would be displayed under the search box but related search suggestions to the left.
This would mean that the suggestions would be displayed differently for each of four search terms conducted by participants.  Displaying the suggestions in two different locations on the screen could have confused the participant,
and confounded the experimental process. It seemed better to display all suggestions in the same location: just below the search box, to maximise consistency.
\item In terms of the number of suggestions to display, the decision was somewhat hampered by
not having access to the vast resources held by widely-used search engines, which could have helped to identify the one most helpful suggestion.
Only offering one suggestion could mean  one meaningless suggestion was offered and the most potentially useful suggestion ommitted. Hence eight of each suggestion type were offered (following Bing's lead), hoping that this would maximise the chances of a meaningful alternative being offered without unduly confusing the participants or producing too much clutter.
\end{itemize}

\section{Configuring SCAMP}\label{scamp}
This experiment will be referred to as ``the synonym experiment''.
Using SCAMP, the synonym experiment was configured as follows:

\begin{itemize}

\item {\em Experimental Conditions:} SCAMP assumes that the experimental condition in the experiments is the IR system/engine.  Thus the researcher  selected and configured a number of search engines to be used in the experiment. Since SCAMP uses the PuppyIR framework it can access the engines available with the framework such as Bing, Twitter, YouTube, iTunes, Wikipedia, Picassa, Flickr, and Digg. 
For the purpose of this experiment  Bing was used across all conditions to eliminate possible differenced between search engine results:
\begin{itemize}
\item (i) Bing with no search aids, 
\item (ii) Bing with spelling suggestions, 
\item (iii) Bing with related search suggestions, and 
\item (iv) Bing with both spelling and related search suggestions. 
\end{itemize}

\begin{figure}[h]
\begin{center}
\includegraphics[width=8cm]{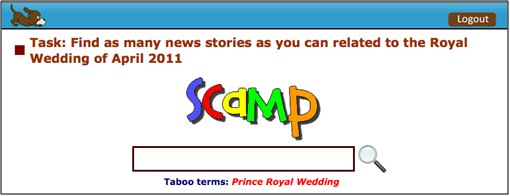}
\caption{Search Instructions}
\label{fig_part}
\end{center}
\end{figure}
\item {\em Assistance Mechanisms:}
 Participants  could be offered spelling suggestions and/or related topic suggestions. Belkin {\em et al.} \cite{Belkin} recommend that suggestions be offered without any effort from the users, as opposed to requiring users to request them so SCAMP simply displayed the suggestions without being prompted.
Spelling suggestions were listed  directly under the search box. A Web Service offered by Ockham (http://code.google.com/p/ockham-spell/) was used
to provide spelling suggestions based on the search term typed in by the user. 
A list of related searches was obtained from the Bing web service. 

\item {\em Search Topics:} The researcher
 specified a  random assignment of topic to assistance mechanism 
 for each participant. A time limit for search tasks was imposed, and
  certain query terms were not allowed (or black listed) as appropriate for each  particular task in order
to ensure that the searcher was forced to come up with synonyms rather than use the most obvious search term.
This is a somewhat contrived way to attempt to replicate the problems people experience in coming up with the optimal search term. 
 Users were told which words they could not use when the topic was introduced and every time they entered a search term (Figure \ref{fig_part})
\begin{figure}[h]
\begin{center}
\includegraphics[width=8cm]{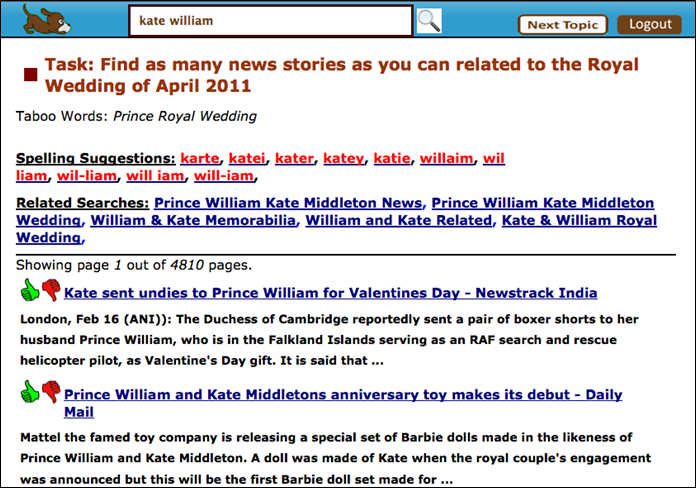}
\caption{Search Results}
\label{fig_part2}
\end{center}
\end{figure}

\item {\em Questionnaires:} Pre/post-experiment  and pre/post-task questionnaires were
specified. 
\begin{figure}[h]
\begin{center}
  \includegraphics[width=8cm]{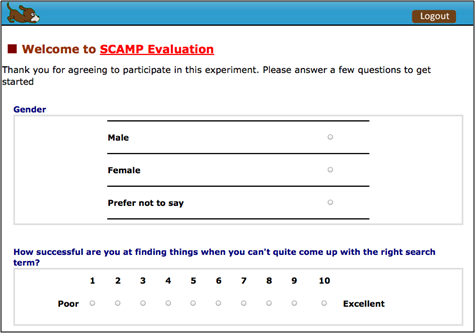}
\caption{Pre-Experiment Questionnaire}
\label{fig_part5}
\end{center}
\end{figure} 

\item {\em Logging:} SCAMP logs relevant user actions. For example, every query term is logged, every document rating, every time a hyperlink is clicked on to view a document (inline) and every question answer (Figure \ref{fig_part3}). This supports later analysis.
\begin{figure}[h]
\begin{center}
\includegraphics[width=8cm]{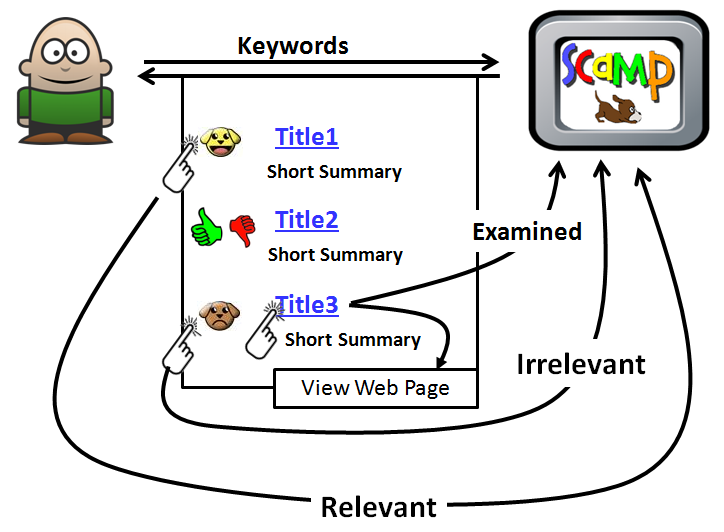}
\caption{Rating Relevance}
\label{fig_part3}
\end{center}
\end{figure}

\item {\em Experimental Tasks}: The tasks were designed to determine how easily users could find information when certain query terms were prevented.  Participants were required to find relevant documents without using taboo terms. For example, if the topic was ``[country name] independence'' then the participants were not allowed to use the terms ``[country name], [country name - ish] independence,  independent''. The restrictions on query terms increased the difficulty of the topics, requiring participants to use synonyms and related terms to find relevant documents.\\

Four search tasks were created based on high profile news stories (in the country in which the study was undertaken).
This search task falls into the first of Bystr\"{o}m and J\"{a}rvelin's \cite{Bystrom} complexity levels: where tasks are completely determinable. In this level some of the tasks may be vague, which was implemented by disallowing obvious search terms. \\

The searchers were asked to find news items about
the Occupy Protest, the	 Royal Wedding, Scottish Independence and the England Riots.
These news items were all prominently reported in the last quarter of 2011 or the first quarter of 2012, so participants are likely to have been aware of them. 
The search tasks were randomly assigned using a latin square rotation to the different experimental conditions. For each search task/topic, taboo search terms were specified.\\

For each task  two web resources were consulted to determine how many synonyms and related terms were available for each of the search tasks (www.lexfn.com; poets.notredame.ac.jp). This ensured that the task was not too difficult: several linked terms and potential synonyms were available. 
Topics with over 50 related terms were chosen. Google Trends was used to ensure that the news story had indeed been a headline story within the previous 12 months so that there was a good chance that participants would be familiar with the general story. \\

\end{itemize}

Figure \ref{fig_part3} demonstrates the mechanics of the participant interface. After the user submits a query a list of results are displayed. Each has a thumbs-up and thumbs-down displayed to the left of the link. The user can indicate that a particular link did indeed match their intent by clicking on the thumbs-up. A smiling dog provides feedback. If the link was not relevant, this can be indicated by clicking on the thumbs-down, and a sad dog provides feedback. (Both actions can be undone by clicking on the icon) If the user wants to view the page, he or she can click on the link, and an embedded window opens just below the link and displays the page. 

Participants could keep searching until their time span had been reached, or continue to the next topic. The experiment concluded with a post-experiment questionnaire, which asked them the question: ``How well do you think you did?'' . 

\section{Results}\label{results}

Forty-Eight participants undertook the web based experiment of which 28 were male, 19 were female, and one declined to give their gender. 
A further 16 participants began the experiment, but did not complete it.
Of the 48, four were eliminated because their data was unusable.  
The experiment was advertised on web based forums and locally. Most participants were from the local 
country, though other participants were based in various countries including South Africa, Germany, Poland, Scotland, England, Holland, Canada and the USA. 

In a pre-experiment questionnaire participants were asked
how well they could generate synonyms to which the mode
was 7 out of 10. 
It seems that most participants felt that they could generate synonyms successfully. 
However, by the end of experiment, 
participants rated how well they did, on average, as 4 out of
10. This suggests that they found the tasks challenging and struggled to find the relevant news items when the most obvious search terms were forbidden. 

Table \ref{table1} summarises the findings of the experiment. 

\begin{table}[h]
\begin{tabular}{|p{2.5 cm}|p{1.25cm}|p{1.5cm}|p{1cm}|p{1cm}|}
\hline
{\bf Condition} & {\bf Avg. Num. Queries} & {\bf Avg. Num. Relevant Results} & {\bf Avg. Time (secs)} \\
\hline
{\bf Bing No Aid}  & 7.35 & 5.27  & 138\\
\hline
{\bf Bing Spell}  & 7.85 & 5.5  & 150\\
\hline
{\bf Bing Related} & 7.93 & 6.16 & 144\\
\hline
{\bf Bing Both} & 7.60 & 6.95 & 144\\
\hline
\end{tabular}
\caption{Usage for each system per task/participant.}
\label{table1}
\end{table}

\subsection{Charts}
SCAMP{\footnote{Charts compliments of Google Charts}} displays some charts to support preliminary analysis. Some are shown here. 
The first thing to determine is whether the participants spent significantly more time using any of the particular engine configurations. 
\begin{figure}[h]
\begin{center}
\includegraphics[width=8cm]{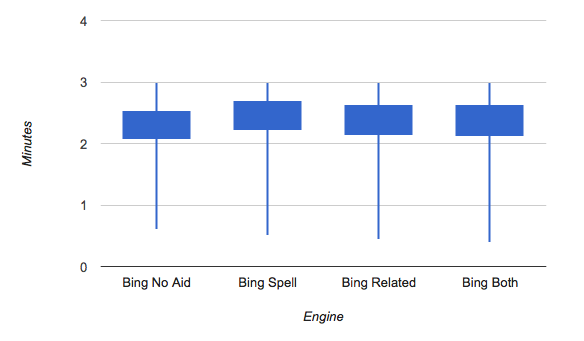}
\caption{Timing per Engine}
\label{fig_part6}
\end{center}
\end{figure}
In Figure \ref{fig_part6} the timing is shown by engine configuration. There is no significant difference in average timings between engines and topics.

\begin{figure}[h]
\begin{center}
\includegraphics[width=8cm]{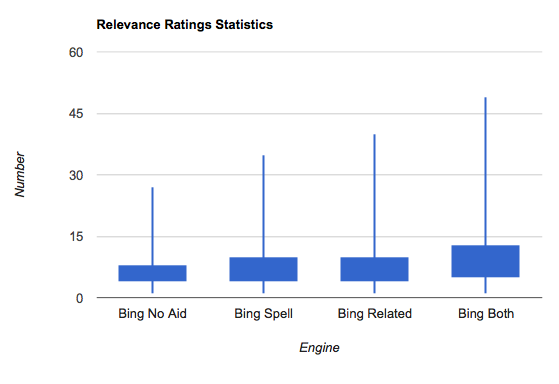}
\caption{Relevance}
\label{fig_part4}
\end{center}
\end{figure}

The hypothesis predicted that people would be able to find more relevant links
if assistance was offered.  The number of relevant links per condition is shown in Figure \ref{fig_part4}. A statistical analysis of the user ratings shows that there is no significant difference between these conditions ($p$ $>$ 0.1 in all cases)

\begin{figure}[h]
\begin{center}
\includegraphics[width=8cm]{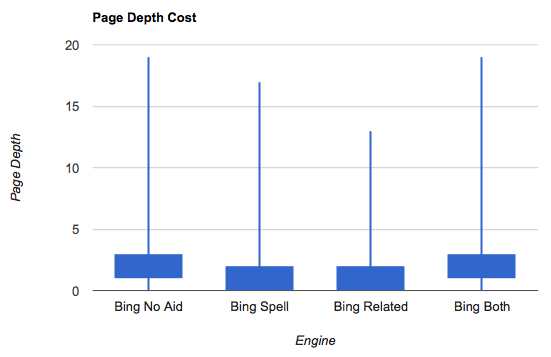}
\caption{Page Depth}
\label{fig_part55}
\end{center}
\end{figure}

Measuring the page depth is a good way to measure effort of querying. In Figure \ref{fig_part55} it is clear that the ``no aid'' and ``both aid'' engines cost more in terms of page depth required to find relevance.  However, this difference is not significant.


\begin{figure}[h]
\begin{center}
\includegraphics[width=8cm]{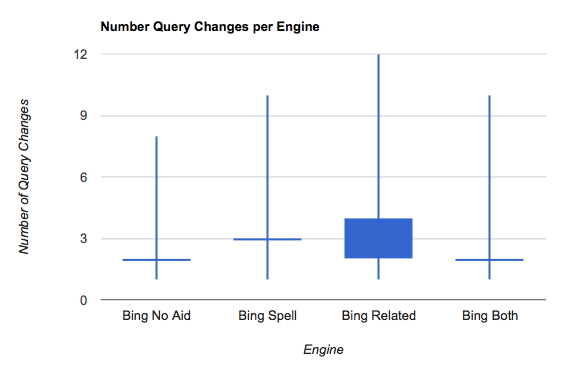}
\caption{Query Changes}
\label{fig_part7}
\end{center}
\end{figure}

In Figure \ref{fig_part7}, one can observe the average query changes per engine. This shows that it is more likely for a participant to change their query should they be presented with related search terms, but not in the other conditions. However, their overall performance was not improved in this condition, which suggests that the related searches were not very useful.

Figures \ref{fig_part55} and \ref{fig_part7} suggest that something unexpected was happening during searching. The participants' comments shed light on the source of the problem.

\subsection{Participant Comments}
Participants did not experience problems using SCAMP's participant interface; they seemed to be able to search and rate results easily. They generally liked the thumbs-up and down buttons, and the immediate feedback when they clicked on these.  However, on a scale of 1 to 10, they rated the search tasks as 5 in terms of how mentally demanding they were. 

Many participants were particularly frustrated by not being able to use the words they wanted to use. Comments such as ``{\em take out the flipping taboo words}'' and ``{\em take away evil forbidden words}'' appeared in response to the post-experiment questionnaire. 

There is also a sense that the suggestions were not used universally by the participants. One said: ``{\em The suggestions for alternate searches and spellings were cluttered and did not catch the eye so were not used.}''. Also: ``{\em i felt it was more my own fault i couldnt come up with better search terms}''. 
It had seemed, during the design phase, that displaying the assistance right under the search box would make them visible and noticeable but it doesn't seem as if everyone noticed and used them. Indeed, when both were offered the interface might well have become too cluttered, thus putting participants off their main task and interfering with search efficiency.

\subsection{Strategies}
The obvious question, when one contemplates this research, is what strategies participants deployed in carrying out the searches.
To support this analysis all queries were classified along the lines of \cite{Rieh06} into one of the following categories:
\begin{enumerate}
\item {\em Initial:}\ \\ The initial query
\item {\em Paging:}\ \\ Progressing to the next page of search results based on the same query.
\item {\em Specialisation:}\ \\ Adding an extra term, eg. ``balcony kiss'' $\rightarrow$ ``balcony kiss kate''
\item {\em Generalisation:}\ \\ Removing a term, eg. ``george square glasgow sit in 2011'' $\rightarrow$ ``george square glasgow''

\item {\em Replacement}: \ \\`Use of synonyms eg. ``l'ecosse independence'' $\rightarrow$ ``alex salmond vote to break away from britain''
\item {\em Taboo Word}:\ \\ eg. ``prince''
\end{enumerate} 

\begin{figure}[h]
\begin{center}
\includegraphics[width=5cm]{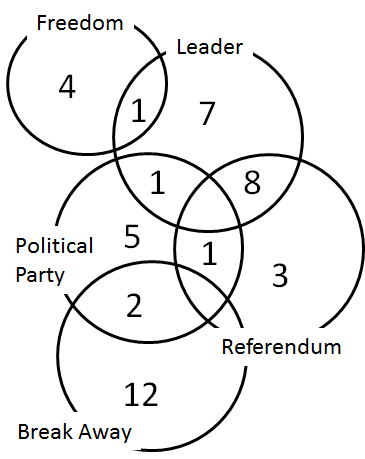}
\caption{Initial Search Terms for Scottish Independence}
\label{venn}
\end{center}
\end{figure}

Figure \ref{venn} shows the initial terms submitted for the topic: Scottish Independence. Four distinct techniques emerge:
\begin{itemize}
\item Hypernyms: Freedom related terms
\item Hyponyms: Breaking away
\item Meronyms: Ruling Political Party \& First Minister
\item Associated Mechanism: Referendum
\end{itemize}
Sensebot {\footnote{www.sensebot.net}} generates 28 terms based on the search ``Scotland Independence". Eight of these are blocked as taboo terms. The participants used nine of the remaining 20. 

Similar tactics were deployed across the other topics but where the topic was related to a specific event searchers would sometimes also provide a date and location. 

\begin{figure}[h]
\begin{center}
\includegraphics[width=8cm]{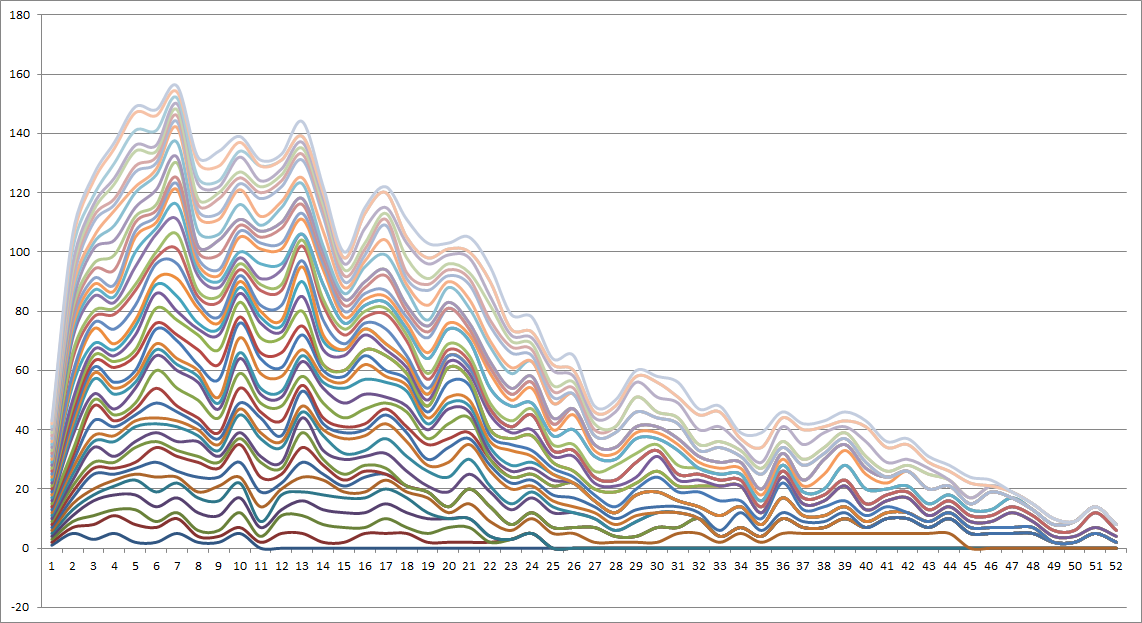}
\caption{Stacked Graph of Participant Strategies}
\label{stacked}
\end{center}
\end{figure}

Figure \ref{stacked} shows a stacked graph of all participants' searches (with the numbers shown next to the categories used to generate the graph). Some persisted much longer than others. The minimum number of interactions with SCAMP was 8, but some participants interacted up to 53 times before timing out. An interesting aspect of this graph is that the activity varies quite a lot at the beginning but the lines plateau towards the end, suggesting a progression to next pages rather than re-formulation.  

The graph in Figure \ref{percs} shows the prevalence of the different strategies across all participants:

\begin{figure}[h]
\begin{center}
\includegraphics[width=8cm]{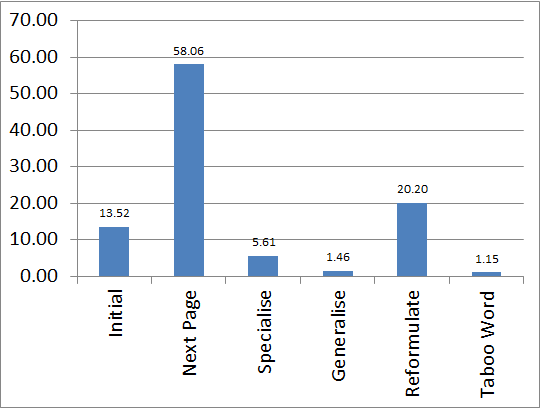}
\caption{Percentages of Categories of Query}
\label{percs}
\end{center}
\end{figure}

Clearly participants proceeded to new pages far more than is usual in searching the web \cite{Spink95}. Spink {\em et al.} found that searchers viewed only 1.7 pages on average, and this effect was consistent across multiple studies. The participants, however, traversed an average of 4.3 pages during their search for each of the four topics. On average, there were only 1.5 query re-formulations per topic. 

A further analysis reveals a correlation of -0.66 between the tendency to proceed to the next page of the results, and the tendency to re-formulate the query. So it seems that those participants who found it hard to come up with synonyms tended to keep searching based on their initial query, and others tried to use synonyms to find what they were looking for. 
The graph also shows that most searchers preferred to traverse successive pages rather than trying to re-formulate the query. 

\begin{figure}[h]
\begin{center}
\includegraphics[width=8cm]{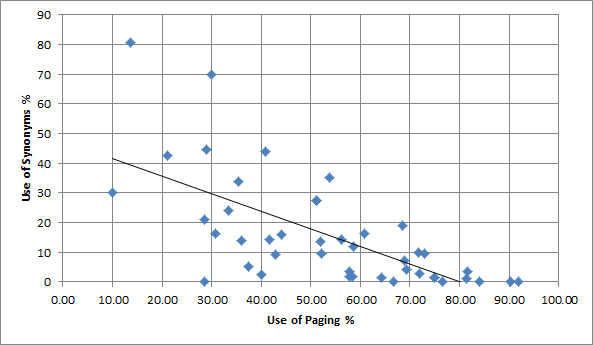}
\caption{Paging Versus Synonym Use}
\label{fig_part8}
\end{center}
\end{figure}

\begin{figure}[h]
\begin{center}
\includegraphics[width=8cm]{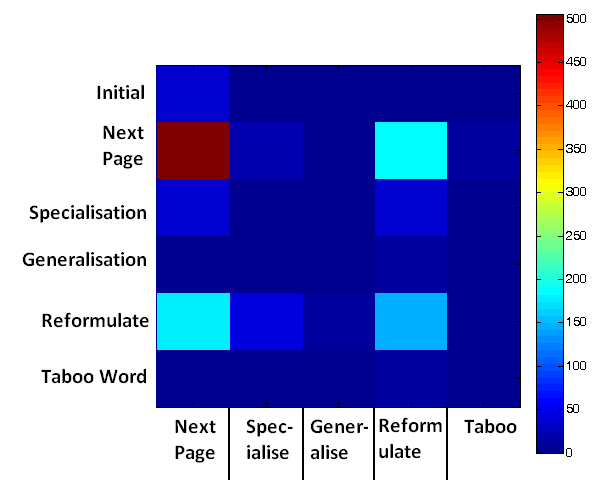}
\caption{Progression From One Query Strategy to Another}
\label{matlab}
\end{center}
\end{figure}

The graph shown in Figure \ref{matlab} plots progression of one query strategy to another across all participants, and this confirms that a majority of searchers seem to choose the strategy of advancing to the next page,
and then stick with it. Very few searchers replaced their query straight after the first one. When they did choose to re-formulate, the most popular action thereafter was also to go straight to the next page, rather than reformulating again, or specialising. 

\begin{figure}[h]
\begin{center}
\includegraphics[width=9cm]{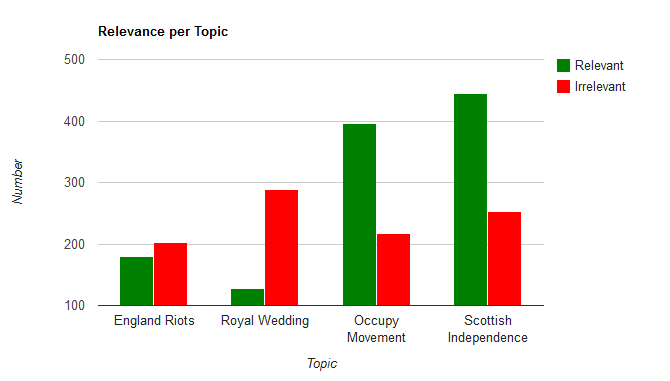}
\caption{Number of Relevant Results Found}
\label{reltopic}
\end{center}
\end{figure}

This analysis concludes by considering the different topics to see whether there were any differences. Figure \ref{reltopic} shows how many results were deemed relevant by the participants.

\begin{figure}[h]
\begin{center}
\includegraphics[width=8cm]{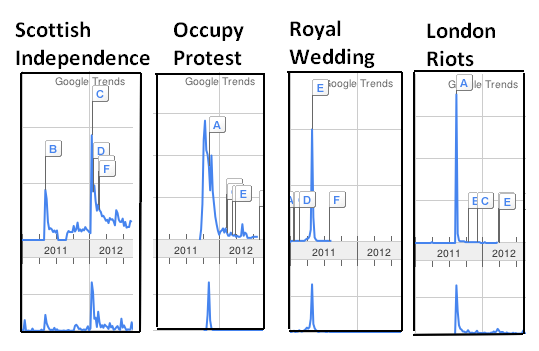}
\caption{Recency of Topics}
\label{reltopic2}
\end{center}
\end{figure}

Scottish Independence and the Occupy Movement  delivered more relevant results than the other two topics. A Google Trends search of 
these topics (Figure \ref{reltopic2}) shows that the two topics which delivered the most relevant results are also the two most recently occurring news events, so that might have played a role. The recency of the news items might have helped searchers to formulate effective query terms.

The number of progressions to next pages and use of re-formulation across
conditions shows an average of progression to new pages of 3.8 for Bing Related as opposed to 4.2, 4.7, and 4.5 for the other conditions. It would be unwise to draw conclusions based on so few participants, especially since the difference is so small. It is possible, however, that the related search
display helped the searcher to come up with query re-formulations, but that the extra clutter when both spelling suggestions and related searches were provided negated this effect. This is, admittedly, pure supposition and would have to be confirmed by a more comprehensive study.

\subsection{Discussion}
To return to the three research questions:

\begin{enumerate}
\item {\em Which aid: spelling or related search suggestions, is most helpful to users who have difficulty finding what they want?}
No differences emerged to show, conclusively, which of these was most helpful. 
The way the assistance was displayed, however, could have caused the interface to become cluttered, and could have negated the potentially positive effects of the assistance. 
Such assistance is routinely offered by search engine query interfaces. They probably have
   strong
  empirical evidence proving  the benefits of offering such assistance but it does not seem to appear in the research literature at present. 
Indeed, the findings appear to confirm those of Park {\em et al.}
\cite{Park} who also found that users did not use the assistance offered to them when searching. 
It would be extremely helpful to know exactly what kinds, and how, to provide assistance in order to be able to conduct information retrieval research effectively. 

\item{\em  Where  on the screen should these suggestions be presented?}
It is clear that offering the suggestions, in the line of sight, needs to be done with care if the display is not to become too cluttered. Offering one spelling suggestion, as the major engines do, is probably the wisest option, but this option was not open to SCAMP, which did not have the vast resources at its disposal to choose the best spelling suggestion.
It was probably unwise, in retrospect, to offer the related work suggestions in the same space. Alternative positions is a fruitful avenue for further research. 

\item{\em  How many suggestions, of each type, should be offered? }
The feedback obtained from the participants showed that fewer suggestions should be offered than
SCAMP currently offers. The number offered by SCAMP clearly cluttered the display. 

\end{enumerate}

The hypothesis was {\em not} supported by the results of the study.
Paired t-tests were carried out to compare all conditions and no significance resulted. 
What does emerge is that people are not particularly apt at 
coming up with alternative search terms when the obvious terms are 
ruled out. This confirms the findings of \cite{Rieh06} about how challenging re-formulation of queries is. It also shows how adaptive people are in finding other ways of reaching their goal, and how they persist with a chosen strategy, even when it does not deliver results.

\section{Conclusion}\label{conc}
The experiment did not show a significant difference between related search suggestions and
 spelling suggestions. This could have been related to the way the information was presented, or it could be that the problem is more fundamental than that. 

An interesting observation was 
that most participants adopted a sub-optimal coping strategy when frustrated in their attempts to use the most obvious search term, probably in an attempt to minimise cognitive effort \cite{Gilovich}.

Future investigations into offering assistance to searchers should focus on the following areas:
\begin{enumerate}
\item Conduct a study using eye tracking equipment to determine how many users actually look at the suggestions.
\item Experiment with different ways of offering suggestions --- as opposed to simply displaying them in a list.
\item Based on these findings, update SCAMP so that suggestions are displayed in the most optimal location, and with as little clutter as possible. 
\item Conduct a larger scale study. The study reported here was a pilot study, and the results serve to highlight the need for further investigation. 
\item Participants were instructed to search for news items. This choice might have confounded searchers, since many felt that the taboo terms made it almost impossible for them to find what they were searching for. It might be worth requiring searchers to find different kinds of items  to see whether that has an impact. (SCAMP can search for images and videos as well as web pages) 
\end{enumerate}

SCAMP is an an Open Source tool which will
be included in the PuppyIR framework that is publicly
available from SourceForge{\footnote{http://sourceforge.net/projects/puppyir/}}.
We would like to get feedback from other researchers so that we can improve SCAMP
as a tool to support researchers investigating search engine technologies. 

\section*{Acknowledgements}
I would like to thank Karen Renaud for reading an earlier draft of this paper and for allowing me to benefit from her expertise in writing about experiments.

\bibliographystyle{apalike}

 \bibliography{example}

\begin{thebibliography}{}

\bibitem[Anick, 2003]{Anick03}
Anick, P. (2003).
\newblock Using terminological feedback for web search refinement: a log-based
  study.
\newblock In {\em Proceedings of the 26th annual international ACM SIGIR
  conference on Research and development in informaion retrieval}, SIGIR '03,
  pages 88--95, New York, NY, USA. ACM.

\bibitem[Belkin et~al., 2001]{Belkin}
Belkin, N.~J., Cool, C., Kelly, D., Lin, S.-J., Park, S.~Y., Perez-Carballo,
  J., and Sikora, C. (2001).
\newblock Iterative exploration, design and evaluation of support for query
  reformulation in interactive information retrieval.
\newblock {\em Information Processing and Management}, 37:403--34.

\bibitem[Broder, 2002]{Broder}
Broder, A. (2002).
\newblock A taxonomy of web search.
\newblock {\em SIGIR Forum}, 36(2).

\bibitem[Brown and McNeill, 1966]{Brown66}
Brown, R. and McNeill, D. (1966).
\newblock The ``tip of the tongue" phenomenon.
\newblock {\em Journal of Verbal Learning and Verbal Behavior}, 5:523--337.

\bibitem[Bystr\"{o}m and J\"{a}rvelin, 1995]{Bystrom}
Bystr\"{o}m, K. and J\"{a}rvelin, K. (1995).
\newblock Task complexity affects information seeking and use.
\newblock {\em Information Processing and Management}, 31(2):191--213.

\bibitem[Fonseca et~al., 2003]{Fonseca03}
Fonseca, B.~M., Golgher, P.~B., De~Moura, E.~S., P\^{o}ssas, B., and Ziviani,
  N. (2003).
\newblock Discovering search engine related queries using association rules.
\newblock {\em J. Web Eng.}, 2(4):215--227.

\bibitem[Gilovich~T, 2002]{Gilovich}
Gilovich~T, Griffin~D, K.~D., editor (2002).
\newblock {\em Heuristics and Biases: The psychology of intuitive judgment}.
\newblock Cambridge University Press, New York.

\bibitem[H{\"o}lscher and Strube, 2000]{Holscher}
H{\"o}lscher, C. and Strube, G. (2000).
\newblock Web search behavior of internet experts and newbies.
\newblock {\em Computer Networks}, 33(1-6):337--346.

\bibitem[Hsieh-Yee, 2001]{Hseih}
Hsieh-Yee, I. (2001).
\newblock Research on web search behavior.
\newblock {\em Library \& Information Science Research}, 23(2):167--185.

\bibitem[Jansen et~al., 1998]{Jansen}
Jansen, Spink, Bateman, and Saracevic (1998).
\newblock Real life information retrieval: {A} study of user queries on the
  web.
\newblock {\em IRFORUM: SIGIR Forum (ACM Special Interest Group on Information
  Retrieval)}, 32.

\bibitem[Jansen and Spink, 2005]{JansenSpink2}
Jansen, B.~J. and Spink, A. (2005).
\newblock An analysis of web searching by european {A}llthe{W}eb.com users.
\newblock {\em Information Processing and Management}, 41(2):361–381.

\bibitem[Jansen and Spink, 2006]{JansenSpink}
Jansen, B.~J. and Spink, A. (2006).
\newblock How are we searching the world wide web? a comparison of nine search
  engine transaction logs.
\newblock {\em Information Processing and Management}, 42(1).

\bibitem[Jansen et~al., 2000]{Jansen00}
Jansen, B.~J., Spink, A., and Saracevic, T. (2000).
\newblock Real life, real users, and real needs: {A} study and analysis of user
  queries on the web.
\newblock {\em Inf.\ Process.\ \& Management}, 36:207--227.

\bibitem[Kelly et~al., 2009]{Kelly09}
Kelly, D., Gyllstrom, K., and Bailey, E.~W. (2009).
\newblock A comparison of query and term suggestion features for interactive
  searching.
\newblock In {\em Proceedings of the 32nd international ACM SIGIR conference on
  Research and development in information retrieval}, SIGIR '09, pages
  371--378, New York, NY, USA. ACM.

\bibitem[Martzoukou, 2005]{Martzoukou}
Martzoukou, K. (2005).
\newblock A review of web information seeking research: considerations of
  method and foci of interest.
\newblock {\em Information Research}, 10(1).

\bibitem[McCray et~al., 2004]{McCray}
McCray, A.~T., Ide, N.~C., Loane, R.~R., and Tse, T. (2004).
\newblock Strategies for supporting consumer health information seeking.
\newblock In Fieschi, M., Coiera, E., and Li, Y.-C.~J., editors, {\em Medinfo
  2004: Proceedings Of THe 11th World Congress On Medical Informatics}. IOS
  Press.

\bibitem[Park et~al., 2005]{Park}
Park, S., Leeb, J.~H., and Baeb, H.~J. (2005).
\newblock End user searching: A web log analysis of {NAVER}, a korean web
  search engine.
\newblock {\em Library \& Information Science Research}, 27(2):203–221.

\bibitem[Renaud and Azzopardi, 2012]{RenaudG}
Renaud, G.~L. and Azzopardi, L. (2012).
\newblock {SCAMP}: {A} {T}ool for {C}onducting {I}nteractive {I}nformation
  {R}etrieval {E}xperiments.
\newblock In {\em Proceedings Fourth Information Interaction in Context
  Symposium (IIiX) 2012, 21-24 August}, Nijmegen, Netherlands.

\bibitem[Rieh and Xie, 2006]{Rieh06}
Rieh, S.~Y. and Xie, H. (2006).
\newblock Analysis of multiple query reformulations on the web: The interactive
  information retrieval context.
\newblock {\em Information Processing and Management}, 42(3):751–768.

\bibitem[Rose and Levinson, 2004]{Rose}
Rose, D.~E. and Levinson, D. (2004).
\newblock Understanding user goals in web search.

\bibitem[Spink et~al., 1995a]{Ozmultu}
Spink, A., Jansen, B.~J., and Ozmultu, H.~C. (1995a).
\newblock Use of query reformulation and relevance feedback by excite users.
\newblock {\em Internet Research}, 10(4):317 -- 328.

\bibitem[Spink et~al., 1995b]{Spink95}
Spink, A., Jansen, B.~J., Wolfram, D., and Saracevic, T. (1995b).
\newblock From e-sex to e-commerce: Web search changes.
\newblock {\em IEEE Computer}, 35(3):133--135.

\bibitem[Sutcliffe et~al., 2000]{Sutcliffe}
Sutcliffe, A.~G., Ennis, M., and Watkinson, S.~J. (2000).
\newblock Empirical studies of end-user information searching.
\newblock {\em JASIS}, 51(13):1211--1231.

\bibitem[Wolfram et~al., 2001]{SpinkBrief}
Wolfram, D., Spink, A., Jansen, B.~J., and Saracevic, T. (2001).
\newblock Vox populi: The public searching of the web.
\newblock {\em Journal of The American Society for Information Science and
  Technology}, 52(1):1073–1074.

\end{thebibliography}

\end{document}